\documentclass[%
 aip,
 amsmath,amssymb,
 reprint,%
]{revtex4-1}

\usepackage{graphicx}  
\usepackage{subfigure}
\usepackage{multirow}
\usepackage{dsfont}
\usepackage{siunitx}
\usepackage{xcolor}

\usepackage[normalem]{ulem}

\usepackage{fancyhdr}
\usepackage{longtable}
\usepackage{parskip}
\usepackage[T1]{fontenc}
\usepackage{dcolumn}   

\usepackage{bm}        
\usepackage{amsfonts}  
\usepackage{amsmath}   
\usepackage{amssymb}   

\usepackage[colorlinks,citecolor=blue,linkcolor=magenta,urlcolor=blue]{hyperref} 

\DeclareMathOperator*{\argmax}{arg\,max}

\newcommand{\bra}[1]{\left\langle #1 \right|}

\newcommand{\pwisein}{\left\{ \begin{array}{ll}}
\newcommand{\pwiseout}{\end{array}\right.}
\newcommand{\ket}[1]{\left| #1 \right\rangle}

\newcommand{\tr}{\operatorname{tr}}
\newcommand{\cost}{\mathcal{L}}
\newcommand{\reg}{\mathcal{R}}
\newcommand{\HS}{\mathrm{HS}}
\newcommand{\stiefel}{\mathds{K}}
\newcommand{\target}{\mathcal{E}}
\newcommand{\regHS}{\reg_{\text{HS}}}
\newcommand{\regChoi}{\reg_{C}}
\newcommand{\regL}{\reg_{L}}

\newcommand{\kb}[2]{|#1\rangle\langle#2|} 
\newcommand{\id}{\mathds{1}} 

\makeatletter
\def\@email#1#2{%
 \endgroup
 \patchcmd{\titleblock@produce}
  {\frontmatter@RRAPformat}
  {\frontmatter@RRAPformat{\produce@RRAP{*#1\href{mailto:#2}{#2}}}\frontmatter@RRAPformat}
  {}{}
}%
\makeatother

\begin{document}

\title{Regularization of Riemannian optimization: Application to process tomography and quantum machine learning }

\author{Felix Soest}
\affiliation{Institute of Theoretical Physics, TUD Dresden University of Technology, 01062, Dresden, Germany}
\author{Konstantin Beyer}
\affiliation{Department of Physics, Stevens Institute of Technology, Hoboken, New Jersey 07030, USA}
\affiliation{Institute of Theoretical Physics, TUD Dresden University of Technology, 01062, Dresden, Germany}
\author{Walter T. Strunz}
\affiliation{Institute of Theoretical Physics, TUD Dresden University of Technology, 01062, Dresden, Germany}

\begin{abstract}  

Gradient descent algorithms on Riemannian manifolds have recently been proposed as a practical tool for optimizing quantum channels.
We add rank-penalizing regularization terms to the cost functions of these methods (in the spirit of Lasso), driving the optimization towards channels with as few Kraus operators as possible.
For quantum process tomography of rank-deficient channels, regularization both accelerates convergence and improves the fidelity of the solution compared to the unregularized case.
For quantum classification tasks, it reduces the channel rank while maintaining classification accuracy, thereby revealing the minimum rank required by the given input data.
\end{abstract}

\maketitle 
\section{Introduction}

Quantum channels or completely positive and trace preserving (CPT) maps describe all physically valid transformations of an arbitrary input quantum state to a corresponding output quantum state. 
In this sense, quantum channels are the most general framework of how information can be processed. 
Quantum computers, for example, implement specifically tailored state transformations to make use of quantum effects for information processing. 
Noise influences the overall channel in a way which is usually detrimental for useful information processing. Similar problems arise in quantum communication schemes, where the transferred message is directly affected by imperfections of the transmission channel.
Currently, many research efforts focus on the question of how to cope with this noise and how to make applications more resistant against it~\cite{lidarQuantumErrorCorrection2013,campbellRoadsFaulttolerantUniversal2017,preskillQuantumComputingNISQ2018,chatterjeeQuantumErrorCorrection2023,sivakRealtimeQuantumError2023,heussenMeasurementFreeFaultTolerantQuantum2024}. In order to do so, it is often instrumental to identify the influence of the noise, that is, to analyze how the channel which transforms the quantum state can be characterized~\cite{martinisQubitMetrologyBuilding2015,erhardCharacterizingLargescaleQuantum2019,harperEfficientLearningQuantum2020, Nielsen2021gatesettomography}. 
Such a task is usually called quantum channel or process tomography and aims at finding a numerical representation of a channel that reproduces the experimentally obtained data~\cite{chuangPrescriptionExperimentalDetermination1997,poyatosCompleteCharacterizationQuantum1997,darianoQuantumTomographyMeasuring2001}.
Various methods have been proposed, ranging from linear inversion and maximum-likelihood methods~\cite{fiurasekMaximumlikelihoodEstimationQuantum2001,sacchiMaximumlikelihoodReconstructionCompletely2001,rahimi-keshariQuantumProcessTomography2011,anisMaximumlikelihoodCoherentstateQuantum2012,whiteNonMarkovianQuantumProcess2022} over convex optimization~\cite{banchiConvexOptimizationProgrammable2020} and projection techniques~\cite{kneeQuantumProcessTomography2018,surawy-stepneyProjectedLeastSquaresQuantum2022} to machine learning approaches~\cite{palmieriExperimentalNeuralNetwork2020,Torlai2023}.
Experimentally, process tomography has been implemented, for example, in superconducting qubits~\cite{bialczakQuantumProcessTomography2010,chowUniversalQuantumGate2012,rodionovCompressedSensingQuantum2014,krinnerBenchmarkingCoherentErrors2020,goviaBootstrappingQuantumProcess2020,whiteDemonstrationNonMarkovianProcess2020}, optical setups~\cite{obrienQuantumProcessTomography2004,shabaniEfficientMeasurementQuantum2011}, trapped ions~\cite{riebeProcessTomographyIon2006}, and nuclear spins~\cite{weinsteinQuantumProcessTomography2004}.

Recently, a method based on Riemannian gradient descent has been proposed for quantum process tomography~\cite{ahmedqpt}. This approach makes use of the fact that Kraus representations of quantum channels form a Stiefel manifold for which efficient optimization techniques exist~\cite{boumal2023intromanifolds, li2020efficient}.
The parametrization in terms of Kraus operators leads to a valid quantum channel by construction, thereby circumventing problems arising in other tomography methods, where the resulting maps have to be projected to the closest valid quantum state transformation after the optimization~\cite{ahmedqpt}.

In general, a full-rank representation of an $n$-qubit quantum channel requires $4^n$ Kraus operators, leading to an exponential scaling in system size.
However, quantum channels arising from circuits with limited connectivity and depth are typically well approximated by low-rank representations.
Methods based on matrix product state representations of the Choi matrix \cite{Torlai2023} or compression approaches \cite{shabani2011, Flammia_2012,riofrioExperimentalQuantumCompressed2017, teoObjectiveCompressiveQuantum2020} can be used in such a case, to enable tomography of higher-dimensional systems. 
For parametrization on the Stiefel manifold, the maximal rank of the channel can easily be constrained by limiting the number of Kraus operators in the representation. 
While such a lower-rank representation does not circumvent the exponential scaling of the individual Kraus matrices $2^n \times 2^n$, it significantly reduces the number of matrices required.
However, if the actual target channel has a rank smaller than the number of Kraus operators, the gradient descent method has no mechanism to favor solutions with a minimal number of nonzero Kraus operators, due to the non-uniqueness of Kraus representations.

In this paper, we investigate the influence of various regularization terms added to the cost function of the Riemannian optimization with the aim of lowering the number of relevant Kraus operators in the representation. 
More specifically, we analyze the performance of three different regularization schemes based on the Hilbert-Schmidt norm of the involved Kraus operators, the purity of the Choi matrix of the channel, and an \(L_1\)-norm of the Stiefel vector that represents the Kraus decomposition, respectively.
The first two terms are directly motivated by the fact that they penalize channels with high rank and large numbers of nontrivial Kraus operators, respectively.
While Ref.~\cite{ahmedqpt} noted the \(L_1\) term in passing, its behavior and justification were not systematically examined. We provide that analysis and include it here for comparison.
All three terms support a convergence of the optimization toward simple representations of the  channel under tomography, with the Hilbert-Schmidt regularization outperforming the other two terms in most cases.

Quantum process tomography is certainly the prime example for numerical channel optimization. 
However, the framework is applicable to more general settings that do not necessarily require learning the full representation of a quantum channel~\cite{10.21468/SciPostPhys.10.3.079,wiersema_optimizing_2022}. 
Quantum machine learning (QML) problems can be of this form, and we will consider them as a second test example for regularized Riemannian gradient descent algorithms. 

Assume we have a classification problem where classical input data is to be discriminated into various classes. 
While, today, such a problem is typically solved by classical algorithms, it has been shown that quantum circuits can be utilized for classification tasks, too~\cite{farhi_classification_2018,mitaraiQuantumCircuitLearning2018,havlicekSupervisedLearningQuantumenhanced2019a,ahmedComparativeAnalysisNoise2025,dingQuantumMachineLearning2025}. There, the input information is encoded in a quantum state, which is then mapped by a quantum transformation to an output state that can be measured to determine the class the input data belongs to.
The task is then to optimize the mapping such that it correctly classifies the input data.

It is an ongoing debate under which conditions quantum machine learning approaches can actually provide an advantage over classical methods. 
Theoretical improvements have been reported~\cite{huang_power_2021, kubler_inductive_2021, Liu2021}.
Still, it is often questioned whether QML will eventually have a practical impact~\cite{PRXQuantum.3.030101, bowles2024better}. Note that the quest for a quantum advantage is not our concern here.
Instead, we show how the regularized Riemannian optimization can yield insight into the properties of a quantum classification problem.
In particular, the regularization terms considered here can help to understand which effective channel rank is necessary to classify a certain  data set.

The paper is organized as follows. We start with a detailed description of the general setting. We then give a brief overview of Riemannian optimization on the Stiefel manifold and introduce the regularization terms we are going to investigate. 
In Sec.~\ref{sec:tomography} and Sec.~\ref{sec:qml} we apply the optimization with regularization to quantum process tomography and a quantum machine learning problem, showing the influence of the terms in illustrative examples.
In Sec.~\ref{sec:conclusion}, we conclude.

\section{Setting}
\label{sec:setting}
We consider the following general optimization scenario for a quantum channel or CPT map.
The aim is to optimize a channel \(T\) that maps an input state \(\rho_\alpha\) to a state \(T[\rho_\alpha]\). A subsequent measurement by a positive operator valued measure (POVM) with elements \(\{M_\beta\}\) would yield outcome \(\beta\) with probability
\begin{align}
\label{eq:pT}
    p_T(\beta|\alpha) = \tr[M_\beta T[\rho_\alpha]],
\end{align}
conditioned on the choice of input state \(\rho_\alpha\).
The probability distribution \(p_T(\beta|\alpha)\) will be subject to a cost function \(\cost_p\) which specifies the desired properties for the channel. \(\cost_p\) depends on \(T\) through Eq.~\eqref{eq:pT}. The channel \(T\) is then optimized to minimize the cost.

The choice of suitable input states \(\rho_\alpha\) and measurements \(M_\beta\), as well as the  specific form of the cost function \(\cost\), depend on the problem to be solved.

\section{Riemannian optimization of Kraus maps}
In order to numerically optimize the channel \(T\), it needs to be represented in a suitable form.
In this paper, we use Kraus decompositions.
Kraus channels with a fixed number of Kraus operators form a Stiefel manifold, and the optimization of the channel can be done by  Riemannian gradient descent on that manifold~\cite{10.21468/SciPostPhys.10.3.079}.
We will review the framework in the following.
More extensive presentations on that matter can be found in Refs.~\cite{boumal2023intromanifolds, tagare_notes_nodate}.

Consider a channel \(T\) on a \(d\)-dimensional quantum system. 
The channel is represented by \(m\) Kraus operators \(\kappa_k\) that satisfy the condition
\begin{align}\label{eq:tp-condition}
    \sum_{k=1}^m \kappa_k^\dagger \kappa_k = \id_d. 
\end{align}
We can introduce the matrix $\stiefel = [ \kappa_1, ..., \kappa_{m} ]^T \in \mathds{C}^{m d \times d}$, created by stacking the Kraus operators.
This allows us to formulate Eq.~\eqref{eq:tp-condition} as
\begin{align}
    \stiefel^\dagger \stiefel =  \id_d,
\end{align}
which is the defining equation of the Stiefel manifold
\begin{equation}
\label{eq:stiefel}
    \mathrm{St}(m d, d) = \{ \stiefel \in \mathds{C}^{m d \times d} : \stiefel^\dagger \stiefel = \id_d \}.
\end{equation}
We can now use a method from optimization on smooth manifolds, namely Riemannian gradient descent (RGD), to optimize a smooth cost function $\mathcal{L}(\stiefel)$~\cite{10.21468/SciPostPhys.10.3.079}.
A normal gradient descent method on the matrix \(\stiefel\) would in general lead out of the manifold defined by Eq.~\eqref{eq:stiefel}.
Therefore, RGD makes use of a retraction $R_\stiefel$, a mapping from the manifold's tangent bundle to the manifold, to bring \(\stiefel\) back to the manifold after a usual gradient descent step:
\begin{align}
    \stiefel' = R_\stiefel (-\epsilon \ \mathrm{grad} \ \mathcal{L}(\stiefel)).
\end{align}
Here, $\epsilon$ is the step size, also known as the learning rate.
For a given manifold, multiple retractions might exist, which can be chosen from freely \cite{boumal2023intromanifolds}.
We use the Cayley transform together with the Sherman-Morrison-Woodbury formula leading to the update rule~\cite{tagare_notes_nodate}
\begin{align}
    \stiefel' = \stiefel - \epsilon U \left(\id + \frac{\epsilon}{2} V^\dagger  U\right)^{-1} V^\dagger \stiefel,
\end{align}
where
\begin{align}
U &= (\Tilde{G}, \stiefel), && V = (\stiefel, - \Tilde{G}), \notag \\
\Tilde{G} &= \frac{G}{|| G ||}, && G_{i,j} = \left(\frac{\partial \mathcal{L}}{\partial \stiefel_{i,j}}\right)^*.
\end{align}
Here, $(\cdot,\cdot)$ represents the row vector of two matrices leading to a matrix of size $(md, 2d)$ and $|| \cdot ||$ the Frobenius norm.
We use $\epsilon = 1$ throughout.

\section{Regularization}
\label{sec:regularization}
In optimization procedures, regularization terms are included in the cost function to favor or penalize solutions with certain properties.
These terms are often used to create simpler or unique solutions. 
The parametrization of the channel \(T\) by Kraus operators leads to an ambiguous solution. 
In fact, any channel has infinitely many different Kraus representations.
We can use the idea of regularization to obtain a solution with a specific form.

For example, let us assume that we have included \(m\) Kraus operators in the Stiefel vector \(\stiefel\), but the optimal solution to the problem is a channel of rank \(r<m\). 
In this case, our model is over-parametrized because the channel could also be represented by only \(r\) different Kraus operators.
If we do not know the rank of the optimal solution, we cannot simply reduce the number of Kraus operators \(m\) in the Stiefel vector \(\stiefel\). However, we can try to penalize models with a large number of finite Kraus operators in \(\stiefel\) and favor those solutions where at least some Kraus operators are close to zero and therefore almost irrelevant for the channel.

The regularization amounts to an additional term \(\reg\) in the cost function. 
This term can depend on the representation \(\stiefel\) of the channel, in contrast to the cost term  \(\cost_p\), which only depends on the channel \(T\) but is otherwise ignorant about its specific Kraus decomposition. 
The complete cost function then reads
\begin{equation}
    \cost = \cost_{p}(T) + \gamma \reg(\stiefel)
\end{equation}
where $\gamma$ is a hyperparameter controlling the regularization strength.

\subsection{Hilbert-Schmidt norm}

The first regularization term we consider is given by the average Hilbert-Schmidt norm of all Kraus operators in the representation,
\begin{align}
    \reg_\HS = \frac{1}{m} \sum_{k=1}^m \sqrt{\tr \kappa^\dagger_k \kappa_k}
\end{align}
This term favors representations with fewer nonzero Kraus operators. Indirectly, this also reduces the rank of the channel.

\subsection{Logarithmic Choi state purity}

As a second regularization term, we consider the logarithmic purity of the channel's Choi state. 
For a channel given by the Kraus operators \(\kappa_k\), the Choi state reads
\begin{align}
    \chi = \sum_k (\kappa_k \otimes \id) \kb{\Phi_+}{\Phi_+} (\kappa_k^\dagger \otimes \id),
\end{align}
with the maximally entangled state
\begin{align}
    \ket{\Phi_+} = \frac{1}{\sqrt{d}}\sum_{j=0}^{d-1} \ket{j} \otimes \ket{j}.
\end{align}
The regularization term is then defined as the negative logarithmic purity of the Choi state $\chi$,
\begin{align}
    \reg_C = -\ln \tr \chi^2.
\end{align}
This term does not explicitly depend on the Kraus representation \(\stiefel\), since it only involves the unique Choi state of the channel. 
However, the purity term favors Choi states of low rank and therefore also Kraus representations with only a few independent Kraus operators.

\subsection{\(L_1\)-norm}
As a third term, we look at the \(L_1\)-norm of the Kraus vector. 
\begin{align}
    \reg_L = | \stiefel |_1 = \max_j \sum_i |\stiefel_{ij}|
\end{align}
This term is inspired by classical Lasso regularization~\cite{lasso} and has been used in Ref.~\cite{ahmedqpt} in a quantum process tomography context. However, there, the authors neither give a profound motivation for the term nor do they analyze its influence on the performance. We therefore include it here to fill this gap. The term can indeed improve the performance, it is however in general outperformed by the two previous choices.

\section{Quantum process tomography}
\label{sec:tomography}
As a first example, let us apply the regularized Riemannian optimization to quantum process tomography (QPT), which is a special case of the general channel optimization setting outlined in Sec.~\ref{sec:setting}.
The aim of such a procedure is to model a quantum channel \(T\) which reproduces the experimentally obtained data for an unknown quantum channel \(\target\). To do so, the input states \(\rho_\alpha\), used in the experiment, form a set of states that span the whole state space. The measurement \(\{M_\beta\}\), performed after the application of the channel, is informationally complete~\cite{bengtssonGeometryQuantumStates2006}. 
This guarantees that there is a unique channel \(T\) whose statistics [see Eq.~\eqref{eq:pT}] reproduces the one of the experimental channel \(\target\).
It is assumed here that the input states \(\rho_\alpha\) and the POVM elements \(M_\beta\) are perfectly known. Strictly speaking, this is never the case in a real experiment. However, this problem is independent of the question we are investigating in this paper. 

Crucially, the probability distribution \(p_m(\beta|\alpha )\) measured in the experiment is in general only an estimate of the true distribution due to the limited amount of measurement data. 
It has been shown that QPT methods nevertheless work in this regime, especially if the channel is not of full rank~\cite{Torlai2023,ahmedqpt}.
However, gradient descent optimization techniques on sparse data are prone to overfitting.
We will show that the regularization terms can help reduce this behavior. 
In order to distinguish this effect from the regularization of the number of relevant Kraus operators in the Stiefel vector \(\stiefel\), we will first look into the case of perfect infinite measurement data (see Sec.~\ref{sec:infinite}), and only later turn toward the experimentally more relevant case of limited measurement data.

In order to optimize the channel \(T\) with the Riemannian gradient descent method, we need a suitable cost function.
A standard cost function for the discrepancy between the measured probability distribution \(p_m\) and the modelled one \(p_T\) is given by the Kullback-Leibler divergence, 
\begin{align}
    \cost_p = \sum_{\alpha} p_0(\alpha) \sum_\beta p_m(\beta | \alpha) \ln \frac{p_m(\beta | \alpha)}{p_T(\beta | \alpha)},
\end{align}
which we will use throughout this section. Here, $p_0(\alpha)$ is the prior distribution over the input states $\rho_\alpha$, which we will assume to be uniform.

We want to investigate the influence of the regularization terms \(\reg\) introduced in Sec.~\ref{sec:regularization}.
These terms are motivated by the idea to minimize the number of relevant Kraus operators in the channel representation \(\stiefel\).
It can therefore be expected that the regularization works in particular for non-full-rank channels. 
A realistic quantum channel always has full rank. 
However, in many cases of practical relevance, only a few Kraus operators with significant magnitude are needed to well approximate the channel.
This will in particular be true for channels that emerge from quantum circuits with limited depth and connectivity as they appear, e.g., in gate-based quantum computers.
Therefore, in the following investigations, we will mainly focus on channels with intermediate rank and analyze how the performance of the regularization depends on the rank.

The overall procedure is as follows. We randomly sample \(n\)-qubit channels \(\target\) of fixed rank and calculate the probability distribution \(p_m(\beta|\alpha)\) that would be measured in an experiment. See App.~\ref{app:channel-init} for details on the sampling of the channels.
As input states \(\rho_\alpha\), we choose all combinations of eigenstates of the Pauli operators, i.e., each input state is of the form 
\begin{align}
    \rho_\alpha = \Pi_{i_1} \otimes \ldots \otimes \Pi_{i_n},
\end{align}
where \(\Pi_{i_j}\) is one of the six eigenstates of the Pauli operators \(\sigma_x\), \(\sigma_y\), \(\sigma_z\) on the \(j\)th qubit.
Up to a normalizing prefactor, these states also form an informationally complete POVM which defines our measurement operators \(M_\beta\).

We then initialize \(T\) in a random channel and use the Riemannian gradient descent method to fit \(T\) to the true channel \(\target\). 
The figure of merit will in general be the infidelity 
\begin{align}
    1 - F(\chi_T,\chi_\target) = 1 - \tr\sqrt{\sqrt{\chi_T}\chi_\target \sqrt{\chi_T}}
\end{align}
between the Choi states of the estimated channel \(T\) and the target channel \(\target\).

\subsection{Infinite shots}
\label{sec:infinite}
We start with the ideal case of infinite measurement data.
Here, the experimentally measured probability distribution \(p_m(\beta|\alpha)\) matches exactly the true statistics of the unknown target channel \(\target\). 
The more realistic case of limited measurement data follows below.

\begin{figure}[h]
    \centering
    \includegraphics[width=\columnwidth]{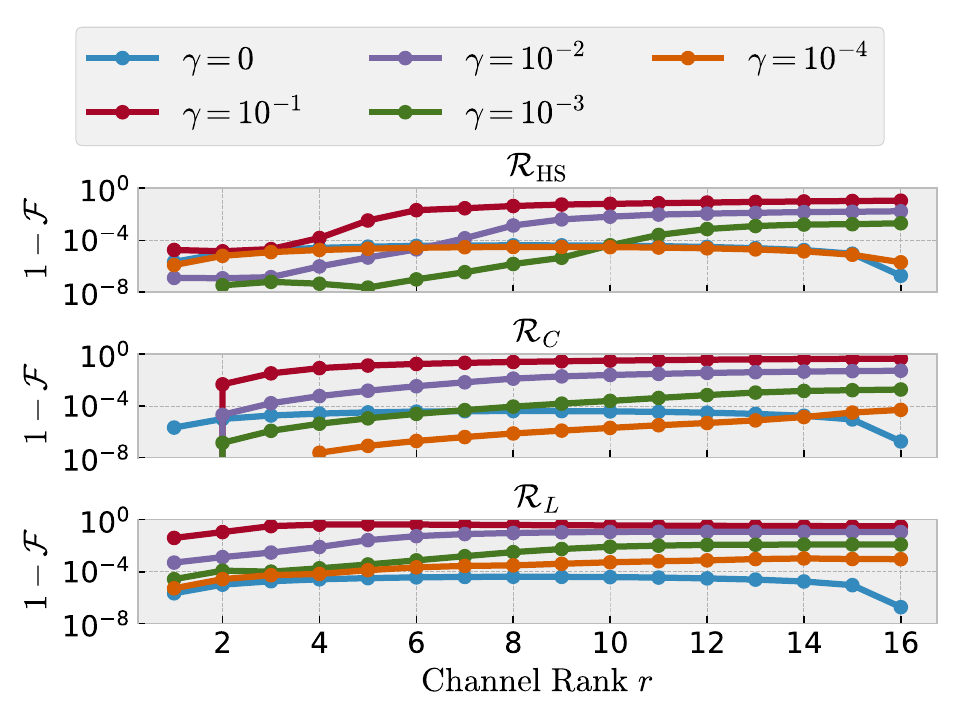}
    \caption{We plot the mean infidelity as a function of the rank of the target channel \(\target\), trained with the regularization terms \(\reg\) after \(10^5\) epochs of training. 
    For each rank, we sample 300 channels. In the infinite shot case considered here, both $\reg_\HS$ and $\reg_C$ provide an improvement over the unregularized case for many target channel ranks and choices of $\gamma$. Thus, these terms can enhance the convergence properties of the optimization. By contrast, the term \(\regL\) leads to a disadvantage for any finite regularization strength \(\gamma\).}
    \label{fig:final_fidelity}
\end{figure}

We consider the tomography of quantum processes for $n=2$ qubits. 
Without prior knowledge of the target quantum channels \(\target\), $m = 4^{n} = 16$ Kraus operators have to be included in the Stiefel vector \(\stiefel\) to be able to cover two-qubit channels up to the maximum rank. 
In Fig.~\ref{fig:final_fidelity}, we plot the impact of the various regularization terms \(\reg\) and strengths \(\gamma\) for randomly sampled target channels with different ranks \(r\).
We see that for a suitably chosen regularization strength \(\gamma=\) \(10^{-3}\), the term \(\regHS\) leads to significant advantages up to a rank \(r=9\). The term \(\regChoi\) is advantageous even up to rank \(r=14\) for \(\gamma = \) \(10^{-4}\). 
The regularization term \(\regL\), based on the \(L_1\)-norm of the Stiefel vector \(\stiefel\), is of no help in the infinite shot regime and the unregularized case performs better on average for all channel ranks.


\begin{figure}[h]
    \centering
    \includegraphics[width=\columnwidth]{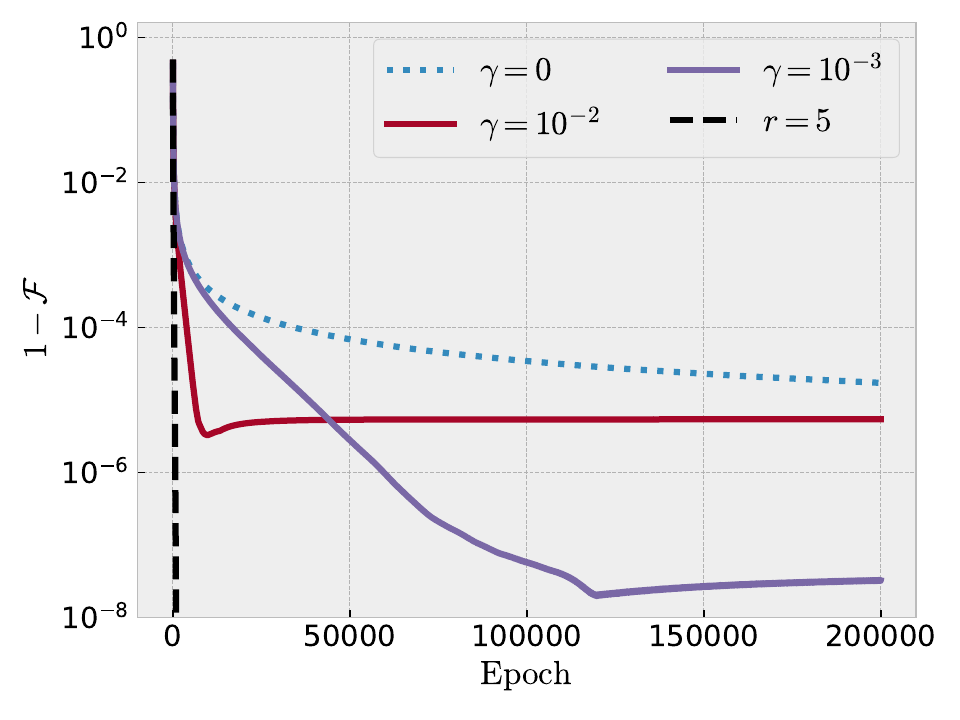}
    \caption{We plot the mean infidelity of 300 randomly sampled target channels with rank $r=5$ in the infinite shot regime. The channel \(T\) is initialized as a random full-rank channel. Using the regularization term $\reg_\HS$, we compare multiple values of $\gamma$. Both non-zero values of $\gamma$ lead to faster convergence compared to the unregularized case \(\gamma = 0\).
    However, if the rank of the target channel \(\target\) is known and the model \(T\) is constraint to it, the convergence is even better~(black dashed line).}
    \label{fig:inf-shots-hs}
\end{figure}

The advantage of the regularization disappears when the true rank of the channel is known.
In that case, one can set the number of Kraus operators $m$ to the channel rank \(r\) and the convergence of the model to the target channel is much faster.
This is illustrated in Fig.~\ref{fig:inf-shots-hs}, where we plot the average training histories of channels with true rank \(r=5\).
The regularization term leads to a lower infidelity on optimization models with $m=16$ Kraus operators in \(\stiefel\).
However, these models are outperformed by models where one sets $m = r = 5$.

\subsection{Finite shots}
We now turn to the case of finite measurements, where each input state $\rho_\alpha$ is prepared $s$ times and the evolved state is measured using the POVM \(\{M_\beta\}\).
The measurement process is simulated by drawing $s$ times from a multinomial distribution with probability $p_\target (\beta | \alpha) = \tr[M_\beta \target[\rho_\alpha]]$ for each input state $\rho_\alpha$.

\begin{figure}[h]
    \centering
    \includegraphics[width=\columnwidth]{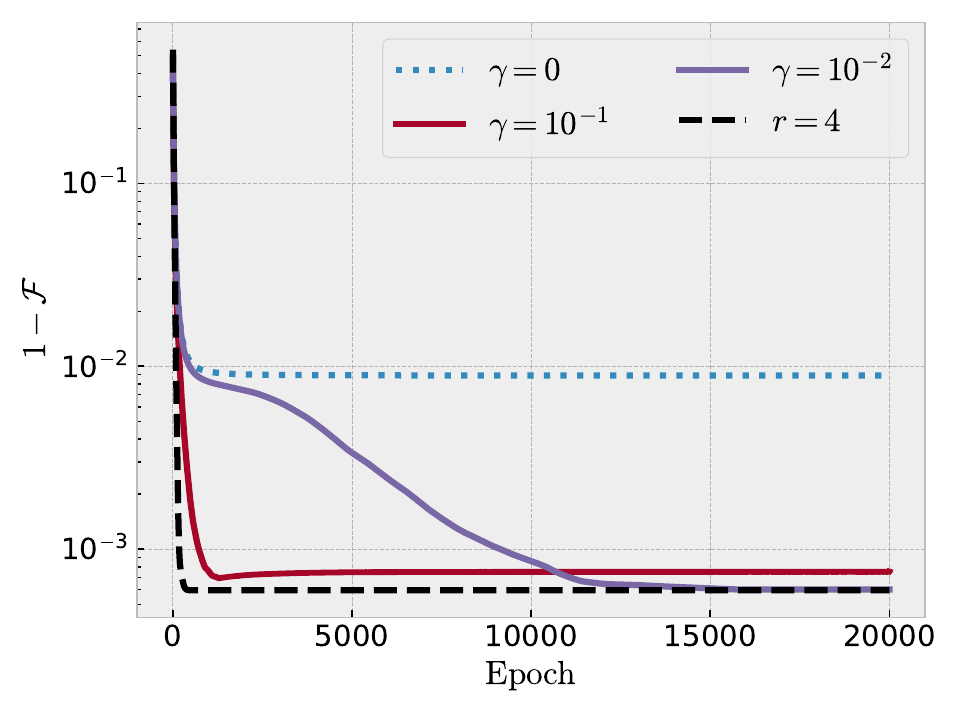}
    \caption{We plot the mean infidelity of 300 target channels with rank $r=4$ for various values of $\gamma$ using $\reg_\HS$. The larger value $\gamma = 10^{-1}$ leads to faster convergence but does not reach the minimum infidelity. The optimal value $\gamma = 10^{-2}$ requires more epochs to converge but reaches the minimum that would also be obtained by a model with \(m=r=4\) Kraus operators (black dashed line). Here, $s=10^4$  measurements are sampled for each input state $\rho_\alpha$.}
    \label{fig:hs-rank4-finite}
\end{figure}

We start the finite shot case by examining the effect of the Hilbert-Schmidt regularization term \(\regHS\) on channels of true rank \(r=4\).
Such channels might arise from a four-qubit unitary process after tracing out two of the four qubits.
In the infinite shot case, we had seen that the optimization converges much faster and to a tighter value, if the true rank of the target channel is known and the number of Kraus operators in \(\stiefel\) are chosen accordingly (cf.~black dashed line in Fig.~\ref{fig:inf-shots-hs}).
As for the convergence speed, this also applies to the case of finite shots. However, as shown in Fig.~\ref{fig:hs-rank4-finite}, with a suitable regularization strength \(\gamma\), the optimization of a full rank model (\(m=16\) Kraus operators in \(\stiefel\)) can reach the same mean infidelity \(1-\mathcal{F}\).
Crucially, the unregularized optimization converges to a much higher value. 
The figure also shows that stronger regularization can lead to faster convergence,  but this comes at the cost of poorer average infidelity.

\begin{figure}[h]
    \centering
    \includegraphics[width=\columnwidth]{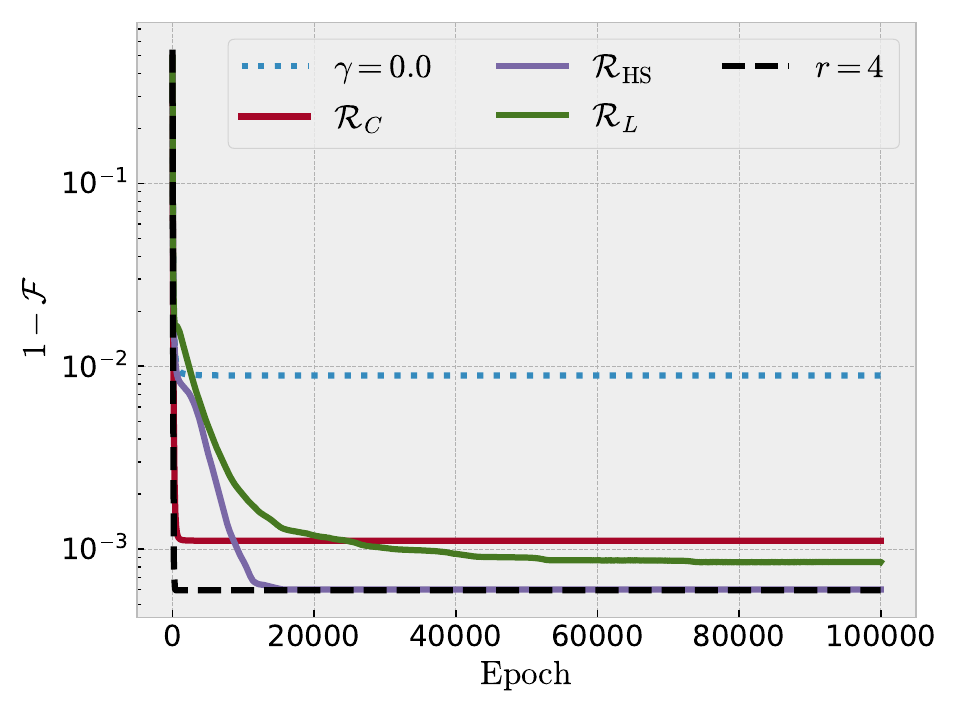}
    \caption{We plot the mean infidelity of models trained with different regularization terms $\reg$. For each curve, 200 target channels are sampled. We perform a grid search over five values of $\gamma$, plotting the value with minimum mean infidelity for each regularization term after \(10^5\) epochs. The performance of models trained with the correct target channel rank $r=4$ is plotted in black for comparison. In the finite shot case with number of shots $s = 10^4$ , all regularization terms reach a significantly lower infidelity for some choice of $\gamma$ than the unregularized case $\gamma = 0$. However, only $\reg_\HS$ reaches the mean infidelity of models where the target channel rank is known in advance.}
    \label{fig:pair_plot}
\end{figure}

The optimal \(\gamma\) with the best mean infidelity in Fig.~\ref{fig:hs-rank4-finite} has been found by a simple grid search over five values of $\gamma \in \{10^{-1},10^{-2},10^{-3},10^{-4},10^{-5}\}$.
We perform the same search for the other two terms \(\regChoi\) and \(\regL\), and plot the infidelities for the respective best \(\gamma\) (with the lowest mean infidelity after \(10^{5}\) epochs) as a function of the training epoch in Fig.~\ref{fig:pair_plot}.
Again, for comparison, we also plot the mean infidelity of a model with \(m=r=4\) Kraus operators in \(\stiefel\). 
Interestingly, in contrast to the infinite shot case, now also the term \(\regL\), based on the \(L_1\)-norm, provides an advantage over the unregularized case. 
However, this term as well as the Choi term \(\regChoi\) cannot compete with the Hilbert-Schmidt term \(\regHS\) in the long run.
On the other hand, for a small number of epochs (up to 8000 for the given example) the term \(\regChoi\) leads to the best results as it causes the fastest drop of infidelity in the beginning of the optimization.

The example shows that suitably chosen regularization terms in the cost function can be advantageous for Riemannian gradient descent approaches to quantum process tomography.
However, the specific choice of a term may depend on whether the aim is to achieve rapid convergence of the algorithm or the highest possible fidelity with the target channel.

\subsection{Optimization of the hyperparameter \(\gamma\)} \label{gamma-optimization}

Up until now, we have compared the impact of the regularization terms by examining the mean infidelity of the estimated channels $T$ and target channels $\target$. In particular, the optimal \(\gamma\) for a specific term was determined from a minimization of the mean infidelity. 

In an experimental setting, the infidelity is, however, not available as the target channel \(\target\) is unknown.
We thus need a different method to optimize the hyperparameter $\gamma$.
Following Ref.~\cite{Torlai2023}, an often-used technique in machine learning to optimize hyperparameters splits the data set into training, validation and test sets.
The machine learning model is trained on the training data, parameter optimization is done using the validation set and its performance is tested on the unseen data in the test set.
Similarly, we can split our measurement results into a training and a validation set to compare the performance of models with different values of $\gamma$.
We forgo using a test set in this setup as the real target channels are known in our benchmarking situation. Therefore we can calculate the fidelity and evaluate the performance of the models directly.
We use an 80/20 train/validation split throughout all models trained in this paper.
To that end, we independently draw the training and validation sets from a multinomial distribution $p(\beta | \alpha)$ for each input state $\rho_\alpha$, splitting the number of shots $s$ per input state.

In order to examine the performance of the regularization terms in an experimental setting where the infidelity is inaccessible, we perform a grid search over various values of $\gamma$ as follows.
We choose a list of 10 values (see App.~\ref{app:grid_search_values}) for the regularization hyperparameter \(\gamma\).
For each $\gamma$, the models are trained on the training set.
Then we choose the optimal $\gamma^*$ by choosing the value with the lowest cost \(\cost\) on the validation set.
As we assume no knowledge of the target channels, we use full rank models to approximate the targets, setting $m=d^2$.

\begin{figure}[h]
    \centering
    \includegraphics[width=\columnwidth]{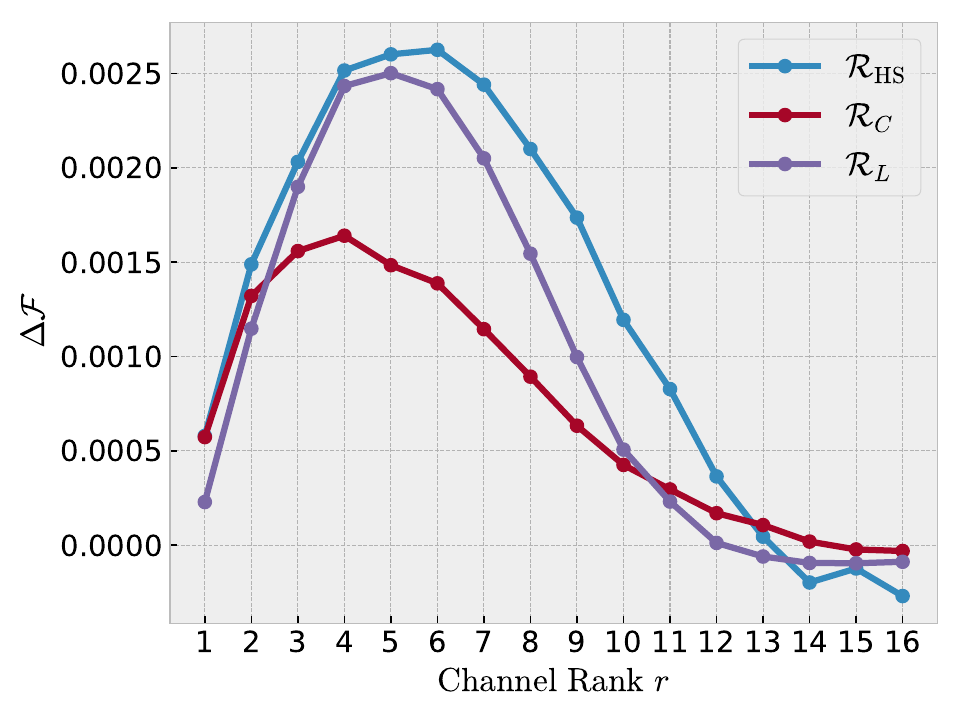}
    \caption{We plot the mean difference $\Delta \mathcal{F} = \mathcal{F}_{\gamma^*} - \mathcal{F}_0$ between the fidelity with optimized regularization strength $\gamma^*$ and the fidelity of unregularized models. We consider this the experimentally relevant case where the fidelity is unknown and thus cannot be used to find the optimal $\gamma$. Instead, $\gamma^*$ is optimized using a grid search over various $\gamma$, choosing the value with the lowest test set cost \(\cost\). All regularization terms provide an improvement over the unregularized case for some target channel ranks. Each model is trained for \(10^5\) epochs with $s = 10^5$.}
    \label{fig:delta_fidelity}
\end{figure}

To benchmark this method, we then look at the difference $\Delta \mathcal{F} = \mathcal{F}_{\gamma^*} - \mathcal{F}_0$ between the fidelity of models with regularization strength $\gamma^*$ and models with no regularization term.
We plot the results for varying target channel rank in Fig.~\ref{fig:delta_fidelity}.
For target channel ranks smaller than twelve, the grid search optimization of the regularization strength \(\gamma\) yields better results on average than training channels without such a term.
This is in agreement with the fact that the regularization mainly works in scenarios where it can be assumed that the target channel is not of full rank.
For the two-qubit case \(d=4\), the method performs particularly well for channels of ranks around \(r=5\). The largest advantages are obtained with the Hilbert-Schmidt term \(\regHS\). However, for ranks beyond \(r=12\) the Choi term \(\regChoi\) performs better.
We stress that this possibility to optimize the hyperparameter \(\gamma\) based on experimentally accessible data makes the regularization a useful improvement of the Riemannian gradient descent method for quantum process tomography. 

The examples in this section show how regularization terms can improve the performance of Riemannian approaches to quantum process tomography. The method works best for rank-deficient channels. Note that a generic
channel is of full rank. However, in many practically relevant settings such as quantum computers, the process is close to a low-rank channel.
In this case, the regularization helps to suppress insignificant Kraus operators without fixing the rank of the optimized channel beforehand.

\section{Quantum classification}\label{sec:qml}
\label{sec:qml}

Let us turn toward another example that fits into the framework of quantum channel optimization.
A typical problem in quantum machine learning is the classification of classical data by means of a quantum mapping. Such a task is similar to the quantum process tomography setting we examined above.
Instead of learning a channel \(T\) that reproduces a probability distribution $p_m(\beta | \alpha)$ given by measurement results of a tomography experiment, the aim here is to optimize a quantum transformation such that it learns a function $y=f(x)$ from a finite data set.
This set  \(\{(x_i,y_i)\}\) includes inputs $x_i$ sampled from some distribution $p(x)$ as well as their corresponding classes labelled by $y_i$.
In order to demonstrate the impact of regularization terms added to the cost function of such a classification problem, we consider two different toy problems.
Firstly, we use the Iris data set \cite{iris-anderson, iris-fisher} which comprises measurements of three species of the iris plant.
Additionally, we consider the Wine data set \cite{wine-dataset}, resulting from chemical analysis of different Italian wines.
This data set is preprocessed using principal component analysis to reduce the input dimension and make it compatible with our setting (see App.~\ref{app:qml-preprocessing}).
Both data sets have three target classes.

Each classical data vector $x \in \mathds{R}^N$ is encoded into a quantum state \(\ket{\psi_x}\). 
The choice of a suitable encoding is often crucial for the performance of a quantum machine learning task, both qualitatively and quantitatively~\cite{PhysRevLett.122.040504, PhysRevA.103.032430}.
However, as we are not so much interested in the overall performance of the classification here but merely in the impact of the regularization, we do not optimize the encoding step in this paper.
Instead, we choose a simple dense angle encoding which can consistently encode the data of both data sets. 
The classical data \(x\) is represented by an \(\lceil N / 2 \rceil\)-qubit pure quantum state of the form~\cite{larose_robust_2020}
\begin{align}
    \ket{\psi_x} = \bigotimes_{i=1}^{\lceil N / 2 \rceil} \cos \frac{\pi}{2} x_{2i -1} \ket{0} + e^{i 2 \pi x_{2i}} \sin \frac{\pi}{2} x_{2i -1} \ket{1},
\end{align}
where \(N\) is the total number of classical features.
The encoded data is mapped by the channel \(T\) and subsequently measured.
Contrary to the process tomography case, the POVM \(\{M_\beta\}\) is given by projectors \(\{\kb{\beta}{\beta}\}\) onto the computational basis. 
Thus, the measured data amounts to a probability distribution \(p(\beta|x)\). 
The desired label \(y\) of the class is then given by the outcome \(\beta\) with the largest probability, i.e.,
\begin{align}
\label{eq:classification}
    y &= f_T(x) = \argmax_\beta p(\beta|x) \notag\\
    &= \argmax_\beta \bra{\beta} T[\kb{\psi_x}{\psi_x}] \ket{\beta}.
\end{align}
We can then use the Riemannian gradient descent method in order to optimize \(T\) such that its statistics yield a good approximation of the desired function, i.e., \(f_T(x) \approx f(x)\).

The channel \(T\) determines the conditional probability \(p(\beta|x_i)\) for each input sample. A typical cost function for our problem is then given by the cross entropy
\begin{align}
    \cost = -\sum_i \ln p(\beta=y_i | x_i)  + \gamma \reg,
\end{align}
where \(\reg\) is again one of the three regularization terms defined in Sec.~\ref{sec:regularization}, and \(y_i\) is the correct classification of input \(x_i\) as given by the training data set.
$\gamma$ is a hyperparameter controlling the strength of the regularization.

We split the data into a training and a test set, optimize the channel \(T\) with the Riemannian gradient descent method on the training data, and apply the result to the test data in order to see how the classifier performs. 
The Iris data set consists of 150 data points with \(N=4\) features which are encoded in a two-qubit input state.
The data belongs to three different classes, i.e., after applying the channel \(T\) to the input states, only three of the four projectors \(\kb{\beta}{\beta}\) are considered to determine the classification outcome in Eq.~\eqref{eq:classification}.
The objects in the Wine data set have 13 features. In a classical pre-processing, we reduce this number to \(N=8\) and encode them in a three-qubit input state (see App.~\ref{app:qml-preprocessing}). 
The total number of 178 data points in this set belong to three different classes. 

Unlike in process tomography, the input states \(\ket{\psi_x}\) generally do not span the whole state space, nor is the measurement informationally complete. 
Thus, in general, the optimal channel \(T\) will not be unique. 
In particular, optimal \(T\) can differ in rank. 
We expect the regularization terms to steer the optimization toward a channel which solves the problem with as few Kraus operators as possible. 
Importantly, this regularization must not degrade the accuracy of the classification itself. 

\begin{figure}[h]
    \centering
    \includegraphics[width=\columnwidth]{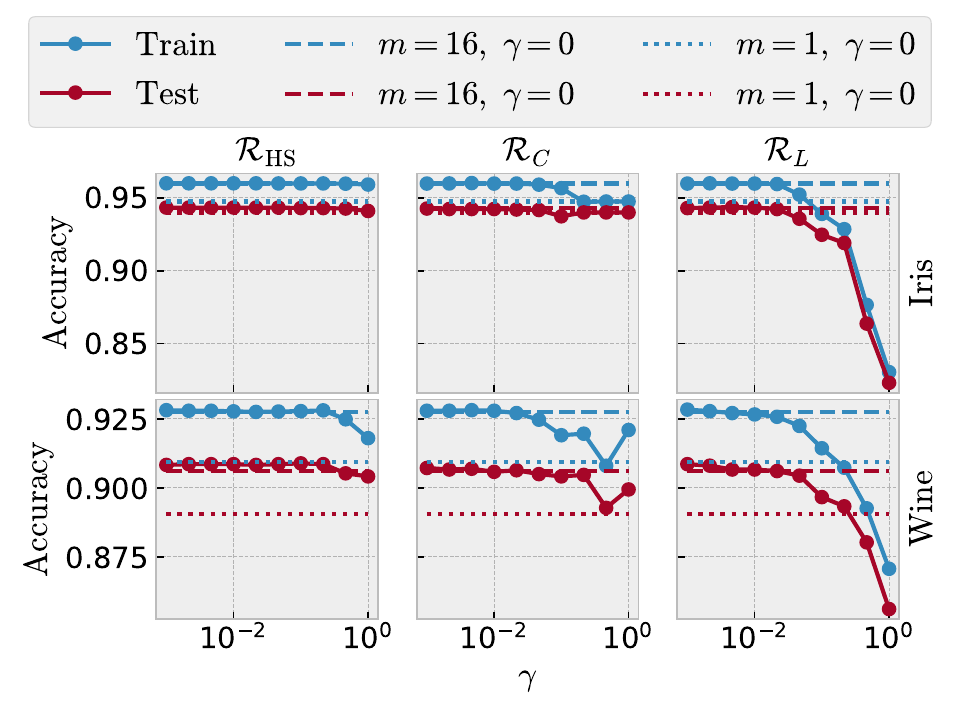}
    \caption{We plot the average accuracy of the classification on training and test data as a function of the regularization strength \(\gamma\).
    For small \(\gamma\), the accuracy is independent of the regularization, thus the additional term \(\reg\) in the cost function does not compromise the accuracy of the classification.
    For comparison, we plot the unregularized case (dashed line). 
    The dotted lines correspond to a unitary model \(m=1\), which performs significantly worse than the general model with \(m=16\) Kraus operators.
    All plotted values are averages over 100 random splits of training and test data after 1500 (Iris data set) or 750 (Wine data set) epochs of training. Each optimization is initialized in a randomly sampled unitary channel. 
    }
    \label{fig:accuracy-qml}
\end{figure}

In order to see that this is indeed the case, we first plot the accuracy of the classification as a function of the regularization strength in Fig.~\ref{fig:accuracy-qml}.
For both example data sets, we see that for sufficiently small regularization strength \(\gamma\), the accuracy both on the training and on the test set is independent of the regularization. For comparison, we plot the accuracy of the unregularized case \(\gamma=0\) (dashed lines).
We also plot the accuracy that can be reached by a unitary model (\(m=1\) Kraus operator, dotted lines), which is significantly worse than the general case of a model with \(m=16\) Kraus operators.
This shows that for a fixed dimension of the quantum system, a non-unitary transformation can in general be a better classifier than a unitary circuit. 
Clearly, according to Stinespring's dilation theorem, the same result could be obtained unitarily if sufficiently many ancilla qubits were added~\cite{stinespring}. 
However, as we will see shortly, the regularized quantum channel model allows us to determine the minimal rank needed to accurately classify the data encoded in a quantum system of given dimension. 
This information can yield insight into the question of what complexity in a quantum circuit is actually needed to classify certain data sets.

\begin{figure}[h]
    \centering
    \includegraphics[width=0.48\textwidth]{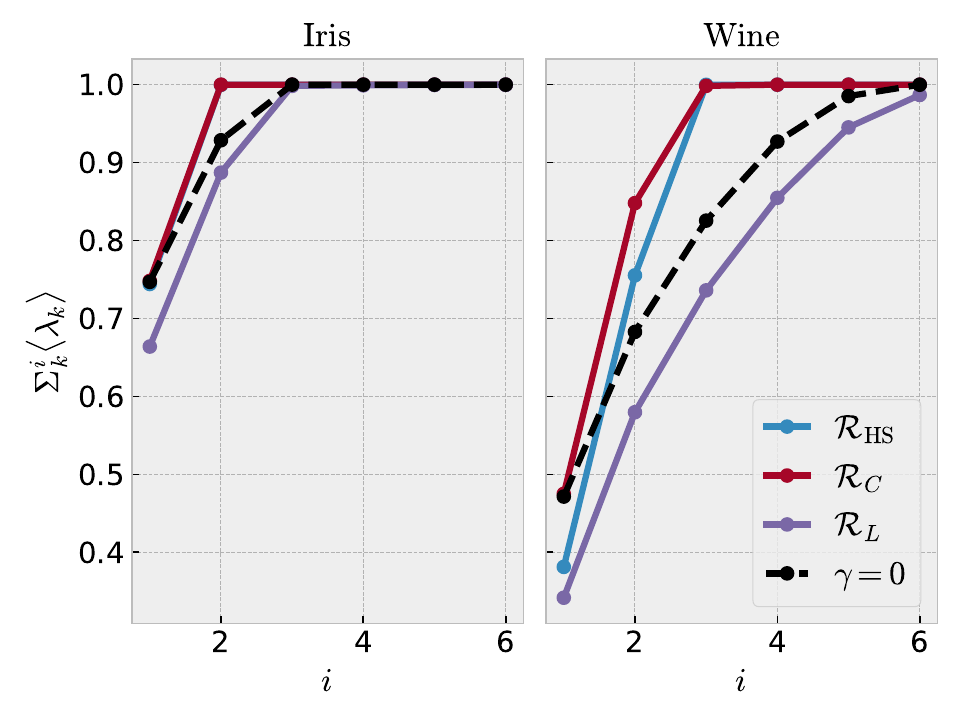}
    \caption{We plot the sum over the \(i\) largest eigenvalues of the channel's Choi state \(\chi\), averaged over 100 random training/test splits and initializations. The regularization terms \(\regChoi\) and \(\regHS\) reduce the rank of the optimized channel \(T\), particularly visible for the Wine data set. The unregularized case (black dashed line) converges to channels of rank six. For the regularized models, already the first three eigenvalues sum to unity, i.e., the channel can be represented by only three Kraus operators. For the Iris data set, the terms \(\regHS\) and \(\regChoi\) reduce the rank from three to two. The term \(\regL\) shows the opposite behavior and is therefore not helpful. For the regularization with \(\regHS\) we use \(\gamma = 0.22\), for \(\regChoi\) and \(\regL\) we have \(\gamma = 0.02\). These values lie in the plateau regions of Fig.~\ref{fig:accuracy-qml}. Thus, they are chosen so as not to compromise the accuracy of the classification.}
    \label{fig:accumulated-eigenvalues}
\end{figure}

For our toy problems, we find that the channel does not need to be full-rank. Instead, the Iris data can be classified by a channel of rank \(r=2\), while rank \(r=3\) is necessary for the Wine data set. 
This becomes visible in Fig.~\ref{fig:accumulated-eigenvalues}, where we plot the average sum of the \(i\) first eigenvalues of the channels Choi state \(\chi\). 
An optimization without regularization converges on average to a channel of rank \(r=3\) for the Iris data set and to a channel of rank \(r=6\) for the Wine toy problem, i.e., only the three (six) largest eigenvalues have a significant magnitude.
However, with the regularization terms \(\regChoi\) and \(\regHS\), only two nonzero eigenvalues remain in the Iris case and three in the Wine case. 
The regularization strength \(\gamma\) was chosen to be in the saturated regions of Fig.~\ref{fig:accuracy-qml}. 
Thus, this regularization decreases the rank of the channel without compromising its classification accuracy. 
The third regularization term \(\regL\) is of no help in this scenario, as it tends to increase the rank of the channel \(T\). This is consistent with the findings for the infinite shot case of quantum process tomography in Sec.~\ref{sec:infinite}.

The examples discussed here highlight the significant role a regularization term can play for the numerical optimization of classifying quantum channels. Remarkably, the classification accuracy is invariant for a wide range of the regularization strength \(\gamma\), as seen in Fig.~\ref{fig:accuracy-qml}. This can be attributed to the fact that the optimal channel for the classification problem is not unique, unlike in a quantum process tomography task. The regularization terms used here lift this ambiguity and favor a low-rank solution. The formalism is flexible and could in the same way include terms for other desired features of the channel, for example a particularly large overlap with a predefined channel that is easy to implement experimentally. 
Additionally, the regularization could include terms that bias the solution according to some classical knowledge we might have about the input data.

\section{Conclusion}\label{sec:conclusion}
In this paper, we show that adding regularization terms to Riemannian gradient descent improves quantum channel optimization when the target channel is rank-deficient.
In quantum process tomography, the method leads to improvements for the accuracy of the optimization of channels of unknown rank. 
The regularization term in the cost function favors solutions with fewer non-zero Kraus operators in the Stiefel vector \(\stiefel\) and supports the convergence of the algorithm in particular for low-rank channels.
Of the three terms considered, the Hilbert-Schmidt term \(\regHS\) performs best in most cases. 
This becomes particularly visible for experimentally relevant scenarios of finite measurement data.

The method sensitively depends on the chosen regularization strength \(\gamma\). 
We discuss how this hyperparameter can be optimized when the target channel is not known in advance, as is the case for quantum process tomography. 
By splitting the measured data into training and validation sets, suitable values of \(\gamma\) can be determined, that outperform the unregularized case \(\gamma = 0\) for most ranks \(r\) of the target channel.
It has to be emphasized, though, that when the target channel rank is known, the regularization terms fail to provide an advantage.
In this case, one can choose the number of Kraus operators in the Stiefel vector \(\stiefel\) to equal the target channel rank, eliminating the need for regularization.

As a second field of application of regularized Riemannian gradient descent, we investigate a simple quantum classification setting.
Instead of classifying data by parametrized unitary circuits, we use a model of full-rank channels. 
Training such a full quantum channel is more costly than a parametrized unitary circuit, so we do not position this as a route to practical quantum advantage.
Rather, it provides insight in the structure of a quantum circuit needed for the classification of specific data.
In particular, the rank-decreasing form of the regularization terms can help to understand which minimum rank is needed to solve a certain classification problem for input data encoded in a quantum system of fixed dimension.

For both quantum process tomography and quantum machine learning, the computational costs for the method scale exponentially in the dimension of the involved quantum system, rendering the approach infeasible for many experimentally interesting scenarios. 
A natural next step is to combine this regularized Riemannian gradient descent with matrix-product-state representations of the channel~\cite{Torlai2023}, which would remove the exponential scaling and make the method a powerful tool for various applications of channel optimization at experimentally relevant system sizes.

\section*{Code availability}
The code will be made available upon reasonable request.

\section*{Acknowledgment}
We are grateful to Oscar Dahlsten for valuable support and encouragement throughout this project.
We thank Richard Hartmann for helpful advice on the numerical implementation. 
The computations were performed on a Cluster at the Center for Information Services and High Performance Computing (ZIH) at TU Dresden. 
KB acknowledges support by the
NSF under award No.\ 2239498
and the Sloan Foundation under award No.\ G-2023-21102.

\bibliographystyle{quantum}
\bibliography{bibliography}

\appendix
\section{Channel initialization}\label{app:channel-init}
To initialize the optimization algorithm and to sample random quantum channels, we follow the method outlined in Ref.~\cite{mezzadri_how_2007}, which is based on the QR decomposition.
This allows us to generate full-rank quantum channels on a given Stiefel manifold.
In Sec.~\ref{sec:qml}, we initialize the parameters $\stiefel_0$ of the optimization algorithm as a unitary channel, regardless of the number of Kraus operators in the optimization manifold $\mathrm{St}(n_k \cdot d, d)$.
To that end, we draw $n_k$ real numbers $x_i \sim \mathcal{U}(0, 1)$ as well as a unitary $u \in U(d)$ drawn from the Haar measure \cite{mezzadri_how_2007}.
The initial parameters are then given by
\begin{align}
    \stiefel_0 = [ \sqrt{x_1} u, ..., \sqrt{x_{n_k}} u ]^T / \sqrt{\Sigma_{i=1}^{n_k} x_i}.
\end{align}

\section{QML Data preprocessing}\label{app:qml-preprocessing}
Both data sets are accessed through \texttt{scikit-learn} \cite{scikit-learn}.

We encode classical data vectors $x \in \mathds{R}^N$ using $\left\lceil \frac{N}{2} \right\rceil$ qubits and employ dense angle encoding as explained in the main text.
Before the encoding, the classical data is transformed as
\begin{align}
    x' = \frac{x - x_{\min} }{x_{\max} - x_{\min}},
\end{align}
where $x_{\max}, x_{\min}$ are the element-wise maximum or minimum respectively.
They are calculated on the training set only.

The Iris data set consists of four-dimensional classical data.
No further preprocessing is needed, and the data can be encoded using two qubits and dense angle encoding.

The Wine data set however has 13 features.
Here, we reduce the dimensionality of the classification problem by performing a principal component analysis on the training data~\cite{jolliffePrincipalComponentAnalysis2016}.
We choose only the six most important principal components.
This preprocessed data set can now be encoded using three qubits and dense angle encoding.

\section{Grid search values}\label{app:grid_search_values}

In Sec.~\ref{gamma-optimization}, the grid search is performed over the values [0.0, 0.0001, 0.000215, 0.000464, 0.001, 0.002154, 0.004642, 0.01, 0.021544, 0.046416, 0.1], where the best value of $\gamma$ minimizes the test set Kullback-Leibler divergence.

\end{document}